\documentclass[aps,prb,twocolumn,superscriptaddress,showpacs]{revtex4}
\usepackage{graphicx}
\usepackage{dcolumn}
\usepackage{bm}

\begin{document}

\title{Inelastic neutron scattering studies of Crystal Field Levels in PrOs$_4$As$_{12}$}

\author{Songxue~Chi}
\affiliation{Department of Physics and Astronomy, The University of Tennessee,
Knoxville, Tennessee 37996-1200, USA}

\author{Pengcheng~Dai}
\affiliation{Department of Physics and Astronomy, The University of Tennessee,
Knoxville, Tennessee 37996-1200, USA}
\affiliation{Oak Ridge National Laboratory, Oak Ridge, Tennessee 37831, USA}

\author{T.~Barnes}
\affiliation{Department of Physics and Astronomy, The University of Tennessee,
Knoxville, Tennessee 37996-1200, USA}
\affiliation{Oak Ridge National Laboratory, Oak Ridge, Tennessee 37831, USA}

\author{H.~J.~Kang}
\affiliation{NIST Center for Neutron Research, National Institute of Standards
and Technology, Gaithersburg, Maryland 20899-6102, USA}

\author{J.~W.~Lynn}
\affiliation{NIST Center for Neutron Research, National Institute of Standards
and Technology, Gaithersburg, Maryland 20899-6102, USA}

\author{R.~Bewley}
\affiliation{Rutherford Appleton Laboratory, Chilton, Didcot, Oxon. OX11 0QX, UK}

\author{F.~Ye}
\affiliation{Oak Ridge National Laboratory, Oak Ridge, Tennessee 37831, USA}

\author{M.~B.~Maple}
\affiliation{Department of Physics,
University of California at San Diego, La Jolla, California 92093, USA}

\author{Z.~Henkie}
\affiliation{Institute of Low Temperature and Structure Research, Polish Academy of
Science, 50-950 Wroc{\l}aw, Poland}

\author{A.~Pietraszko}
\affiliation{Institute of Low Temperature and Structure Research, Polish Academy of
Science, 50-950 Wroc{\l}aw, Poland}

\date{\today}

\begin{abstract}
We use neutron scattering to study the Pr$^{3+}$
crystalline electric field (CEF) excitations in the filled 
skutterudite PrOs$_4$As$_{12}$. By comparing the observed levels
and their strengths under neutron excitation with the theoretical 
spectrum and neutron excitation intensities, we identify the Pr$^{3+}$ CEF levels, 
and show that the ground state is a magnetic $\Gamma_4^{(2)}$ triplet, and 
the excited states 
$\Gamma_1$,  
$\Gamma_4^{(1)}$
and 
$\Gamma_{23}$ are
at 0.4, 13 and 23~meV, respectively. 
A comparison of the observed CEF levels in PrOs$_4$As$_{12}$
 with the heavy fermion superconductor PrOs$_4$Sb$_{12}$ 
reveals the microscopic origin of the differences in the ground states 
of these two filled skutterudites.
\end{abstract}


\maketitle

\begin{center}
${\bf I.~~INTRODUCTION}$
\end{center}

The Pr-based filled skutterudites (FS) have the formula
PrT$_4$X$_{12}$, where T is one of the transition metals Fe, Ru, or Os,
and X is a pnictogen (P, As, or Sb) \cite{Nolas,Sales,Chakou}.
The notably mounting interests and efforts in the study of 
the FS compounds are motivated by the remarkable diversity 
of their electronic and magnetic ground states, including
multipole ordering \cite{Aoki02a,Sugawara},
small gap insulators \cite{Sekine,Matsunami},
conventional superconductivity \cite{Yogi,Takeda}, unconventional
superconductivity \cite{Bauer02a,Maple01} and magnetic ordering
\cite{Bauer02b,Butch,Yuhasz,Maple06,Adroja}. 
Despite the large differences in their physical properties, these 
compounds are governed by only a few parameters, including   
the interaction between the conduction and the $4f$ shell electrons
(the c-f coupling) and the effect of the crystalline electric field (CEF) potential on
the Pr$^{3+}$ $4f$ electrons 
\cite{Aoki02a,Sugawara,Sekine,Matsunami,Schotte,Bauer02a,Maple01,Bauer02b,Butch,Yuhasz,Maple06,Adroja}.
For example, transport and bulk magnetic measurements on the heavy Fermion superconductor 
PrOs$_4$Sb$_{12}$ suggested either a $\Gamma_1$ singlet ground state or a $\Gamma_3$ nonmagnetic 
doublet ground state \cite{Bauer02a,Maple01}. Inelastic neutron scattering experiments on 
PrOs$_4$Sb$_{12}$ showed that the Pr$^{3+}$ CEF levels include
a $\Gamma_1$ singlet ground state and a low-lying $\Gamma_4^{(2)}$ 
magnetic triplet 
excited state at 0.6~meV \cite{Goremychkin,Kuwahara04,Kuwahara05}. 
This rules out the quadrupolar Kondo effect, which arises only from
a nonmagnetic doublet ground state \cite{Cox}, as the microscopic
origin for the observed heavy-fermion superconductivity. 

The FS compounds belong to the space group Im\={3} \cite{Chakou}. The rare earth atoms
are located at the corners and body-center of the cubic lattice, each 
of which is surrounded by a simple cube of 8 transition metal atoms at the 
8c sites [Fig.~1(a)] and by an icosahedron of 12 pnictogen atoms at the 
24g Wyckoff sites [Fig.~1(c)]. 
Owing to their unique structure, a subtle 
modification on composition can
result in a different CEF scheme and thus a completely different ground state. 
However, a general understanding is desirable as to 
how the compositions influece the CEF levels.
In PrOs$_4$As$_{12}$, in which the pnictogen 
Sb in PrOs$_4$Sb$_{12}$ is replaced by As, 
the material displays quite different correlated electron properties 
\cite{Yuhasz,Maple06}. The temperature dependence of the electrical resistivity 
reveals Kondo lattice behavior, which is not observed in PrOs$_4$Sb$_{12}$ \cite{Aoki02b}. 
Specific heat measurements indicate an enhanced electronic specific heat coefficient of
$\gamma\approx 1$~J/mol~K$^2$ for $T\leq1.6$~K and $0 \leq H\leq 1.25$~T \cite{Yuhasz}.
The compound exhibits several ordered phases at temperatures below 2.3~K and fields below
about 3~T \cite{ho}. The ground state has been determined 
to be antiferromagnetic (AF) by neutron scattering experiments \cite{Maple06}.
A determination of the Pr$^{3+}$ CEF level scheme in PrOs$_4$As$_{12}$ 
and its microscopic origin is crucial for understanding why  
its ground state is different from that in PrOs$_4$Sb$_{12}$.  The outcome will lead to a more
general understanding of how the structures and compositions in Pr-based FSs can influence
their CEF levels and ground states.

\begin{figure}
\includegraphics[width=3.2in]{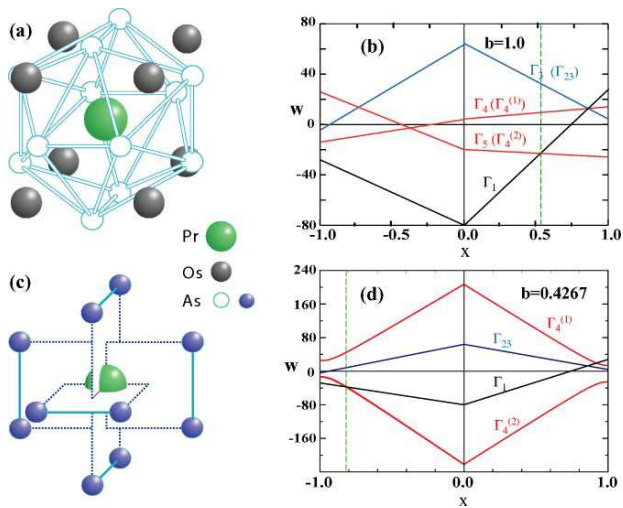}
\caption{\label{fig1} (Color online) 
(a) The cube of 8 Os ions surrounding the central Pr$^{3+}$ ion in PrOs$_4$As$_{12}$.
These give an $O_h$-symmetric CEF.
(b) The corresponding spectrum of $O_h$-symmetry Pr$^{3+}$ CEF levels. 
(black=singlet, blue=doublet, red=triplet). The relative coupling $x$ that gives singlet-triplet
degeneracy is shown by a dashed vertical.
(c) The 12 nearest-neighbor As ions surrounding the central Pr$^{3+}$ in PrOs$_4$As$_{12}$,
giving a reduced symmetry ($T_h$) CEF. 
(d) The corresponding As-only $T_h$-symmetry Pr$^{3+}$ CEF spectrum in PrOs$_4$As$_{12}$.
}
\end{figure}

\begin{center}
${\bf II. ~~EXPERIMENTAL}$
\end{center}

PrOs$_4$As$_{12}$ single crystals were grown using the molten
metal flux method described in Ref.~\cite{Yuhasz} and crushed into fine powder. 
Our neutron scattering experiments were 
carried out on the cold neutron triple-axis spectrometer SPINS at
the NIST Center for Neutron Research (NCNR) and on
the HET chopper spectrometer 
at ISIS (Rutherford Appleton Laboratory), as described previously \cite{Wilson}.
We reference positions in reciprocal space at wave
vector ${\bf Q}=(q_x,q_y,q_z)$ in \AA$^{-1}$ using 
$(H,K,L)$ reciprocal lattice units (r.l.u.) notation,
where $(H,K,L)=(q_xa/2\pi,q_ya/2\pi,q_za/2\pi)$ for
the cubic PrOs${_4}$As$_{12}$ unit cell ($a=8.5319$ \AA) \cite{Yuhasz}.
We used a $^3$He-$^4$He dilution refrigerator for the field-dependent experiments.
The nature of observed CEF excitations were confirmed in a large temperature
(0.08~K-200 K) and magnetic field (0~T-11 T) range.

\begin{figure}
\includegraphics[width=3.2in] {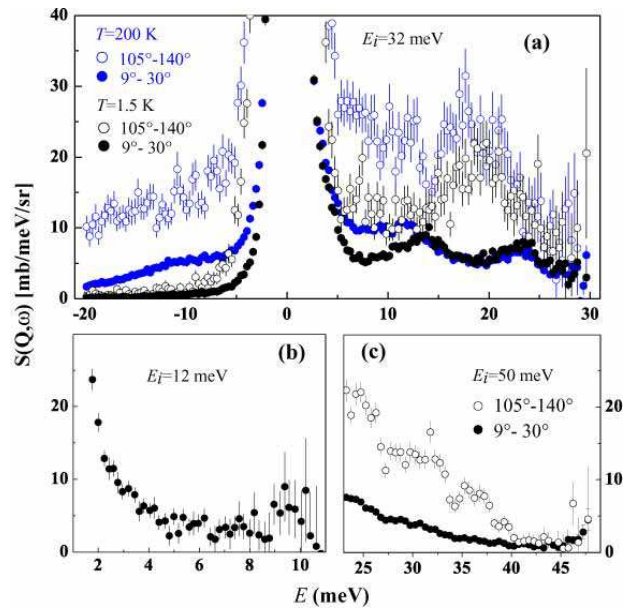}
\caption{\label{fig2} (Color online) (a) Neutron inelastic scattering at
$T=1.5$~K and $T=200$~K with $E_i=32$~meV, integrated  
over scattering angles from 9$^\circ$ to 30$^\circ$ (low-angle detectors) and 
from 105$^\circ$ to 140$^\circ$ (high-angle detectors).
(b) The same at $E_i=12$~meV; (c) $E_i=50$~meV.  The scattering function
$S(Q,\omega)$ was normalized by comparison to a vanadium standard.} 
\end{figure}

\begin{center}
${\bf III. ~~RESULTS~AND~DISCUSSION}$
\end{center}

Figure~2 summarizes the neutron scattering intensity from
PrOs$_4$As$_{12}$ on HET at temperatures between 1.5~K and 200~K. Since the
CEF magnetic scattering decreases with increasing $Q$ 
whereas the intensity of phonons increases with $Q$, 
a comparison of the neutron intensities in the low- and high-angle detectors 
can distinguish between magnetic and phonon scattering. 
Figure~2a shows the scattering function at $T=1.5$~K and $T=200$~K with an incident neutron beam energy of 
$E_i=32$~meV. Comparison of the low- and high-angle data reveals two clear CEF excitations 
at 13~meV and 23~meV, with phonons at $\sim$~20~meV.  Measurements 
with $E_i=12$ and 50~meV showed no evidence of additional 
CEF excitations at energy transfers between 2 and 8~meV or above 25~meV [Figs.~2(b) and 2(c)]. 

To search for CEF excitations at energies below 2~meV, 
we carried out high resolution measurements using SPINS. 
At $T=0.32$~K, energy scans at $Q=(1.2,0,0)$ showed a clear peak at 
0.4~meV; this mode decreases and becomes broader on warming to 
2.5~K and 6~K [Fig.~3(a)]. Figure 3(b) shows that the energy of the $\sim$0.4 meV mode has weak 
${Q}$-dependence and decreases in
intensity with increasing ${Q}$, thus confirming its magnetic nature.
Figure~3(d) reveals that 
the elastic intensity also decreases on warming from 0.08~K to 4~K. This reduction of intensity
in the elastic channel with increasing temperature is also observed in the HET data, evidencing 
the ground state is a magnetic multiplet. 

Figures~4(a)-(d) show the temperature dependence of the low-angle scattering for $E_i=32$~meV. 
The CEF peak intensities do not change significantly with temperature between 1.5~K and 5~K. 
At 50~K the intensity in the elastic channel has undergone a substantial decrease, 
and the 13~meV peak has shifted to 10~meV. On further increasing the temperature to 
100~K and 200~K the intensities at 0~meV, 13~meV and 23~meV continue to decrease, 
whereas the scattering at 10~meV increases. 

The theoretical description of the Pr$^{3+}$ CEF levels in PrOs$_4$As$_{12}$
is complicated by the presence of important contributions from two sets 
of neighboring ions, Os and As. The Pr$^{3+}$ ion in Pr-based FS has a $4f^2$ configuration,
which in Russell-Saunders coupling has a ninefold degenerate $^3$H$_4$ ground state.
This degeneracy is lifted by the CEF interaction, which we assume to be dominated 
by the 12 nearest neighbor pnictogens (As) and the 8 next nearest neighbor Os ions; 
the distances to these ions are $d_{\rm Pr-As} = 3.23~$\AA \ and $d_{\rm Pr-Os} = 3.69~$\AA \ 
respectively. 

\begin{center}
${\bf A. ~~Single-charge~model~with~separate~ions}$
\end{center}

Os ions form a simple cube around the Pr ions and they alone give an $O_h$ symmetric CEF.  
The arrangement of the 12 pnictogens (As) around 
Pr$^{3+}$ forms 3 orthogonally intersected planes where the
As-As bonds are shown as solid lines with length $L$ and 
$W$ is the length of the dashed lines in Fig. 1(c) ($b=W/L=0.4267\neq 1$). 
When the 4 pnictogen
atoms in each of the 3 orthogonally intersected planes form a square, {\it i.e.,}
$b=W/L=1$, the fourfold rotational symmetry is recovered and the point group 
symmetry becomes $O_h$ with the simple cubic CEF potentials [Fig. 1(c)] \cite{LLW}.
This $O_h$ case is treated by Lea, Leask, and Wolf (LLW) \cite{LLW} (see their Fig. 9); we have 
rederived their excitation spectrum as shown in Fig. 1(b). 

Both $O_h$ and $T_h$ CEF interactions split the Pr$^{3+}$ $^3$H$_4$ ground state 
into a singlet, a doublet and two triplets \cite{Takegahara01}.
In the $b\neq 1$ $T_h$-symmetry case, these multiplets are referred to as 
$\Gamma_1$ (a singlet),
$\Gamma_{23}$ (a nonmagnetic doublet), and
$\Gamma_4^{(1)}$ and $\Gamma_4^{(2)}$ (magnetic triplets). These two triplets 
are linear combinations of the $O_h$-symmetry triplets, mixed by the new 
$T_h$ CEF interaction \cite{Takegahara01}; this mixing modifies the excitation spectrum
and leads to $b$-dependent neutron transition intensities.

The $T_h$-symmetry CEF excitation spectrum 
has not been considered in detail in the literature,
and the corresponding neutron transition intensities between $T_h$ CEF levels 
have not been considered at all. To aid in the interpretation of our neutron scattering data we 
carried out these CEF calculations using a point charge model. 
We assumed an expansion of the perturbing CEF potential in spherical harmonics,
\begin{equation}
V(\Omega) = \sum_{\ell = 4,6} g_{\ell} \sum_{m=-\ell}^{\ell} M_{\ell m} Y_{\ell m} (\Omega),
\end{equation}
where the interaction strengths $g_4$ and $g_6$ are treated as
free parameters. The spherical harmonic moments $\{ M_{\ell m}\}$ are determined by the 
positions of the 12 As ions, which we assigned the (scaled) coordinates 
$\vec x =$ 
$(\pm 1, \pm b, 0)$,
$(0,     \pm 1, \pm b)$, 
$(\pm b, 0,     \pm 1)$.
The nonzero independent moments for $\ell = 4,6$ are 
$M_{40} = 21 (1-3b^2+b^4) / 2 \sqrt{\pi} (1+b^2)^2 $,
$M_{60} = 3 \sqrt{13} (2-17b^2+2b^4) / 8 \sqrt{\pi} (1+b^2)^2 $
and
$M_{66} = -15 \sqrt{3003} b^2 (1-b^2)/ 16 \sqrt{\pi} (1+b^2)^3$.  
The nonzero $M_{66}$ for $b\neq 1$ ($T_h$ symmetry) confirms
the presence of the $B^6_t$ terms of 
Takegahara {\it et al.} [Eq.(7) of Ref.\cite{Takegahara01}], 
in addition to the usual $B^4_c$ and $B^6_c$ $O_h$-symmetry 
terms. (Note that the $T_h$-allowed moment $M_{66}$ vanishes at the
$O_h$-symmetry point $b=1$.)
We also confirmed that the other nonzero moments
satisfy the ratios quoted in Eq.(7) of Ref.\cite{Takegahara01}. 
Unlike Takegahara {\it et al.} \cite{Takegahara01}, we do not introduce
a new parameter $y$ for the $T_h$-symmetry terms, 
because they are completely determined by $g_6$ 
and the lattice parameter $b$ in the point charge model. This 
was previously noted by Goremychkin {\it et al.} \cite{Goremychkin}.

Diagonalization of this $T_h$ CEF interaction within the Pr$^{+3}$ $^3$H$_4$ 
nonet gives our results for the spectrum of CEF levels and their associated eigenvectors.
These eigenvectors depend only on the ratio $g_6/g_4$ and the lattice parameter $b$; 
the energies in addition have an arbitrary overall scale.  
Our results for the spectrum for $b=1$ ($O_h$ symmetry) and 
$b = 0.4267$ (PrOs$_4$As$_{12}$ geometry) are shown in Figs. 1(b) and 1(d), 
using LLW normalization conventions \cite{LLW}. (These conventions set
our two Hamiltonian parameters in Eq.(1) to
$g_4 = (968\pi/21)x $ and $g_6 = (-5808\pi/221)(1-|x|)$.)  
Note that the $b=1$ and $b=0.4267$ level schemes differ qualitatively,
which demonstrates the importance of the $T_h$ terms in this problem.   

We find that the $O_h$ singlet and doublet energy eigenvectors
are unmodified by the $T_h$ interaction, consistent with Takegahara {\it et al.} 
\cite{Takegahara01}. The singlet eigenvector (in a $J_z^{tot}$ basis) is 
$|\Psi_1\rangle = \sqrt{7/12} |0\rangle + \sqrt{5/24}(|4\rangle + |-4\rangle)$
and the two doublet states are
$|\Psi_{23a}\rangle = -\sqrt{5/12} |0\rangle + \sqrt{7/24}(|4\rangle + |-4\rangle)$ 
and
$|\Psi_{23b}\rangle = \sqrt{1/2}(|2\rangle + |-2\rangle)$, 
consistent with earlier (numerical) results \cite{Takegahara01,LLW}.
The singlet and doublet energy eigenvalues in our conventions are modified by the 
$T_h$ interaction. In terms of the LLW variable $x$ \cite{LLW} 
and our parameter $b$ they are
$E_1 = -(16/13(1+b^2)^2) (91x(1-3b^2+b^4) -20 (1-|x|)(2-17b^2+2b^4))$
and
$E_{23} = -(16/13(1+b^2)^2) (13x(1-3b^2+b^4) +16  (1-|x|)(2-17b^2+2b^4))$. 
The corresponding analytic results for the two $T_h$ triplet states for general 
$b$ are quite complicated, so we only present numerical results for these states. 

The neutron transition intensities are defined by 
$I_{if} = |\langle f | J_z^{tot} | i \rangle|^2$, 
as introduced by Birgeneau \cite{Birgeneau}.
(There is an implicit sum over initial and final magnetic quantum numbers.)
Our $T_h$-symmetry results for these quantities are shown in Fig. 5(a).  
The values in the limits $x=\pm 1$ 
(no $\ell = 6$ term, hence $O_h$ symmetry) 
implicitly check Birgeneau's numerical $O_h$ results; 
see the off-diagonal entries in his Table~1(e). 
These $O_h$ limits are indicated on the vertical axis of Fig. 5(a).  

Next we compare the observed CEF levels and their neutron excitation intensities 
to the well-known LLW CEF results for $O_h$ symmetry [Fig. 1(b)] and our calculated
CEF predictions for PrOs$_4$As$_{12}$ under $T_h$ symmetry [Fig. 1(d) and Fig. 5(a)]. 
Both $O_h$ and $T_h$ 
CEF spectra have $x$ values that can accommodate a magnetic triplet ground state and 
a nearly degenerate singlet first excited state [vertical lines in Figs. 1(b) and 1(d)].
However, it is evident that the $O_h$ scheme cannot explain the data because 
the observed 0.4~meV transition $\Gamma_{4}^{(2)}\to \Gamma_1$ 
is incorrectly predicted 
to have zero intensity due to the $O_h$ symmetry.
While in the $T_h$ scheme  
the relative neutron excitation strengths of the higher levels (at 13 and 23 meV) 
predicted in Fig. 1(d) seem to be in good agreement with observation at low temperatures,
the As CEF alone predicts an incorrect 
spectrum of levels [Fig.~1(d)], with the triplet $\Gamma_{4}^{(1)}$ being the highest 
excitation. The calculated neutron transition intensity shown in Fig.5(a) can not explain 
the observed intensity at higher temperatures.
As temperature increases, the excited states get populated and the excitations 
start to decrease in intensities. Meanwhile the new excited-state 
transitions start to increase. If $\Gamma_{4}^{(1)}$, instead of $\Gamma_{23}$, is the 
highest level, the intensity at 
12.6~meV would not increase but that at 22.6 meV would, because the
$\Gamma_1$ to $\Gamma_{23}$  transition is not allowed even in $T_h$ symmetry.
Goremychkin {\em et al}. \cite{Goremychkin} showed that the highest level 
in the similar Sb material is the $\Gamma_{23}$ doublet.  

\begin{figure}
\includegraphics[width=3.2in] {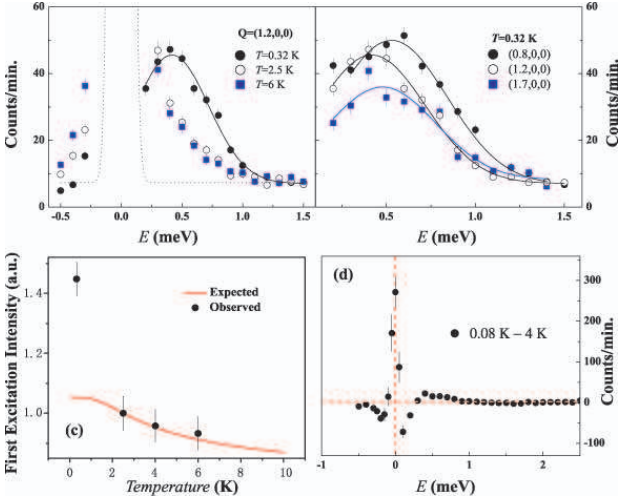}
\caption{\label{fig3} (Color online) 
a) Low energy spectrum of
CEF excitations observed at $T=0.32$, 2.5 and 6.0~K using
the SPINS spectrometer at NCNR.
(b) The wave vector dependence of the excitations at $Q=(0.8,0,0)$, $Q=(1.2,0,0)$, and $Q=(1.7,0,0)$.
(c) The expected and observed temperature dependence of the intensity of the 0.4~meV mode. 
(d) The temperature difference spectrum between 0.08~K and 4~K,
showing clear reduction in magnetic elastic scattering.
}
\end{figure}

\begin{figure}
\includegraphics[width=3.2in] {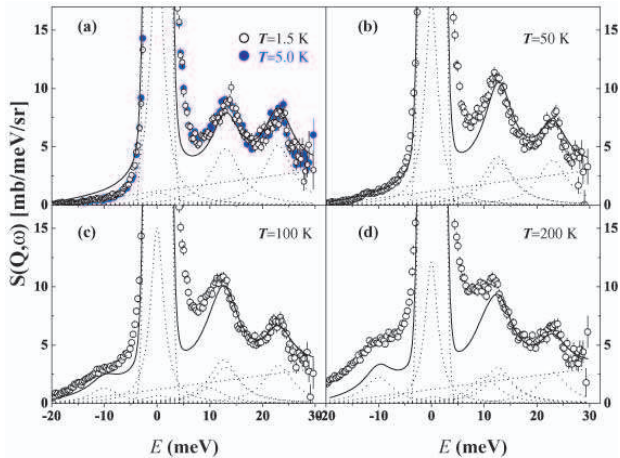}
\caption{\label{fig4} (Color online) 
The temperature dependence of the excitations observed on
HET with $E_i=32$ meV at (a) $T=1.5$~K, 5~K; (b) 50~K;
(c) 100 K, and (d) 200 K.
The lines are theoretical results for neutron excitation
intensities, from the combined Os-As CEF model, with an arbitrary overall scale factor.
}
\end{figure}

\begin{figure}
\includegraphics[width=3.2 in]{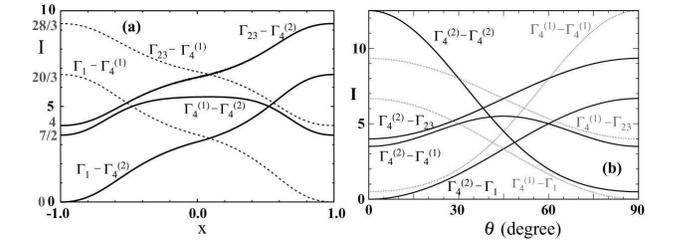}
\caption{\label{fig5}  
(a) The theoretical neutron transition intensity for As CEF alone with b=0.4267.
(b) Neutron excitation intensities predicted by the combined Os-As CEF model.
}
\end{figure}

\begin{center}
${\bf B. ~~Combined~Os-As~CEF~model}$
\end{center}

The twin constraints of having the $\Gamma_{23}$ level at the top of the spectrum {\em and}\
having a large $\Gamma_4^{(2)} \leftrightarrow  \Gamma_1$ 
neutron excitation strength requires both Os and As
terms in the CEF interaction. We therefore introduce a combined Os-As Hamiltonian,
\begin{equation}
H = H(Os) + H(As).
\label{eq:Hfull}
\end{equation}
Although this model nominally has four parameters 
($g_4^{Os}, g_6^{Os}, g_4^{As}$ and $g_6^{As}$), only three are independent; 
$g_4^{Os}$ and $g_4^{As}$ cannot be distinguished because they are summed 
into a single coefficient of the $O_h$-symmetry $\ell=4$ interaction. 
For this reason we introduce combined $O_h$-symmetry Os-As coefficients 
$g_4 = g_4^{Os} + g_4^{As}$ and $g_6 = g_6^{Os} + g_6^{As}$, 
which we normalize according to LLW conventions. 
As the $\ell = 6$ $T_h$-symmetry terms from $H(As)$ in the CEF are proportional
to $g_6^{As}$ alone, the strengths $g_6^{Os}$ and $g_6^{As}$ can
be distinguished. We parametrize these two $\ell=6$ interactions using 
the total $O_h$-symmetry $g_6$ and a $T_h/O_h$ relative strength $r_6$, which
is the ratio of the coefficients of $Y_{62}$ to $Y_{60}$ in the CEF potential.
The energy levels of this Hamiltonian are 
$E_{4(2)} = -6g_4 -  8g_6 - f$,
$E_{1} =    28g_4 - 80g_6$,
$E_{4(1)} = -6g_4 -  8g_6 + f$ and
$E_{23} =    4g_4 + 64g_6$, where
$f = ( (20g_4 + 12 g_6)^2 + 960\, r_6^2 g_6^2)^{1/2}$. 
For $r_6=0$ these reduce to the familiar LLW $O_h$ spectrum.
In the pure As model, $r_6$ is determined by CEF theory if we assume point 
As ions, and is given by $(11\sqrt{105}/4)b^2(1-b^2)/(1+b^2)(1-(17/2)b^2+b^4)$. 
For PrOs$_4$As$_{12}$ we have $b=0.4267$, which gives a rather large 
$r_6=-6.901$. This drives strong level repulsion between the two triplets, 
which explains why the pure As spectrum of Fig.1d 
differs so greatly from the $O_h$ (pure Os) symmetry spectrum of Fig.1b. 

Our experimentally observed CEF levels are close to but not exactly 
consistent with the predictions above of the mixed Os-As model, since the gap ratio 
$(E_{4(1)}-E_{4(2)})/(E_{23}-E_{4(2)}) \approx 0.57$ is slightly {\it below} the 
theoretical lower bound of 7/12. The parameters we estimate from the measured gaps are
$g_4\approx 0.24$~meV and $g_6\approx 0.20$~meV. The value of $r_6$ is not determined by the
measured energies due to the inconsistency mentioned above, 
although $r_6 \lesssim 0.5$ appears plausible.
A more sensitive determination of $r_6$ is possible through the measurement 
of the inelastic neutron excitation intensities we discuss below.

The neutron excitation intensities in this combined Os-As Hamiltonian 
depend only on a single parameter $\theta$, which is the mixing angle of the 
triplet energy eigenvectors when expanded in an $O_h$-symmetry 
$|3\rangle,|3'\rangle$ basis,
\begin{eqnarray}
&
|4(1)\rangle  =
&+ \sin(\theta) 
| 3 \rangle 
+ \cos(\theta) 
| 3' \rangle 
\nonumber
\\
&
|4(2)\rangle  =
&+ \cos(\theta) 
| 3 \rangle 
- \sin(\theta) 
| 3' \rangle.
\label{eq:theta}
\end{eqnarray}
This mixing angle is related to the Hamiltonian parameters by 
$\tan(2\theta) = 2\sqrt{15}\, r_6/(5 (g_4/g_6) + 3)$. 
The singlet and doublet $O_h$ energy eigenvectors are unchanged. 
The nonzero neutron excitation intensities in terms of
$s = \sin(\theta)$ and $c = \cos(\theta)$ are
${\Gamma_4^{(2)}} \leftrightarrow {\Gamma_1}           = (20/3)s^2$,
${\Gamma_4^{(2)}} \leftrightarrow {\Gamma_4^{(1)}}     = 7/2 + 8 c^2 s^2$,
${\Gamma_4^{(2)}} \leftrightarrow {\Gamma_{23}}        = 4 + (16/3)s^2$,
${\Gamma_1}       \leftrightarrow {\Gamma_4^{(1)}}     = (20/3)c^2$,
${\Gamma_4^{(1)}} \leftrightarrow {\Gamma_{23}}        = 28/3 - (16/3)s^2$,
${\Gamma_4^{(2)}} \leftrightarrow {\Gamma_4^{(2)}}     = (25/2)(1-(4/5)s^2)^2$,
and
${\Gamma_4^{(1)}} \leftrightarrow {\Gamma_4^{(1)}}     = (1/2)(1+4s^2)^2$. 
The calculated neutron scattering intensity of different transitions as a function 
of $\theta$ is shown in Fig. 5(b).
We recover the $O_h$-symmetry results of Birgeneau 
(Table~1(e) of Ref.\cite{Birgeneau}) for $s=0,c=1$.

We carried out a least-squares fit of our neutron excitation data at 1.5~K, 
50~K, 100~K and 200~K (Fig. 4) to the theoretical intensities given above, 
which gives an estimate of the triplet mixing angle $\theta$ in PrOs$_4$As$_{12}$, 
\begin{equation}
\theta \approx 22.5^{\circ}.
\label{eq:theta_expt}
\end{equation}

When combined with the values of $g_4$ and $g_6$ from the spectrum,
this $\theta$ corresponds to $r_6 \approx 1.2$. In this fit the relatively isolated 
$\Gamma_4^{(2)} \to \Gamma_4^{(1)}$ peak at 23~meV was used to infer 
the background, 
which was taken to be constant plus linear. The assumed lineshapes 
were Lorenzians with a common linewidth, fixed by the 23~meV peak. 
The calculated intensities of the individual
transitions (dotted lines) and the total intensity (solid lines)
for each temperature are shown in Figs.~4(a)-(d). 
We note that the intensity reduction at 0.4 meV on warming from 0.32 K to
2.5 K is larger than that
expected from the CEF model [Fig. 3(c)], thus suggesting Pr-Pr interactions 
below $T_N$ (=2.3 K) are important.  On the other hand, 
the large difference between the calculated
and expected intensity around 8~meV in
the 200~K data is presumably due to thermally populated phonons (Fig.~2a).


\begin{center}
${\bf C. ~~Field~effect~on~the~CEF~gap}$
\end{center}

\begin{figure}
\includegraphics[width=3.2in]{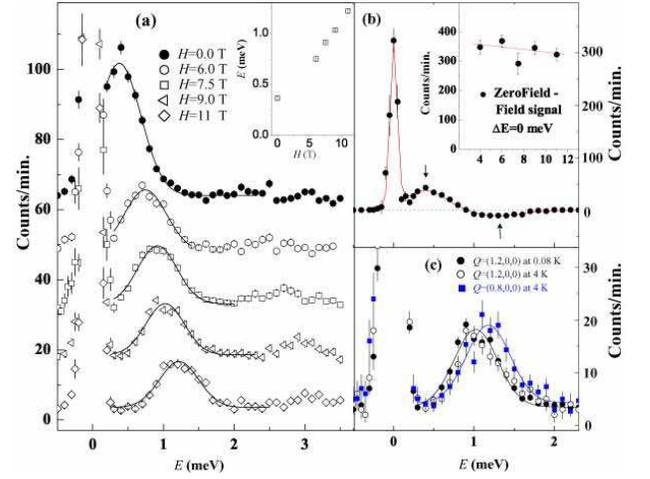}
\caption{\label{fig6} (Color online) (a) 
The magnetic field dependence of the low-lying CEF excitations
at $Q=(1.2,0,0)$ and $T=0.08$~K. The inset shows the field
dependence of the first excited state.
(b) The difference spectrum between 0~T and 11~T at $Q=(1.2,0,0)$ and $T=0.08$~K. 
The effect of an applied field is to suppress intensity at $\hbar\omega=0$~meV, and 
to split the spin-triplet ground state; the latter results in the field dependence 
of the 0.4 meV peak in (a), which may involve an intra-triplet transition. 
The elastic intensity suppression effect essentially disappears for fields above 4~T. 
(c) The field-induced CEF excitation at $\approx 1.1$~meV is weakly 
wave vector dependent, and shows essentially no temperature dependence between 
$T=0.08$~K and 4~K.
}
\end{figure}

If the ground state of PrOs$_4$As$_{12}$ is indeed the $\Gamma_4^{(2)}$ triplet, 
application of a magnetic field should Zeeman split it, resulting in a field dependent energy
gap. 
There should also be a reduction in the intensity of the zero-energy 
$\Gamma_4^{(2)}\to\Gamma_4^{(2)}$ magnetic scattering. 
Figure~6 shows that these expectations are indeed qualitatively satisfied. 
The first excited state at 0.4~meV shifts toward higher energies as the applied field increases.
The field-dependent transition energy is linear only at higher fields (between 6~T and 11~T). The 
drop of intensity in the elastic channel is almost constant for all applied fields, as shown in
the inset of Fig. 6(b). Figure~6(c) shows that the wave vector dependence is also present with applied magnetic field 
($H=9~T$). 
Normally the field-splitting of the ground state multiplet 
would be a very clear test of our $\Gamma_4^{(2)}$ triplet assignment for the ground state. However, 
PrOs$_4$As$_{12}$ is complicated by the near degeneracy of the $\Gamma_4^{(2)}$ and $\Gamma_1$
levels, which mix strongly under an applied field. This results 
in a more complicated spectrum of low-lying states, with several low-field level crossings and neutron scattering intensities that are also modified by their field-induced 
$\Gamma_4^{(2)}$-$\Gamma_1$ mixing.    

\begin{figure}
\includegraphics[width=3.2 in]{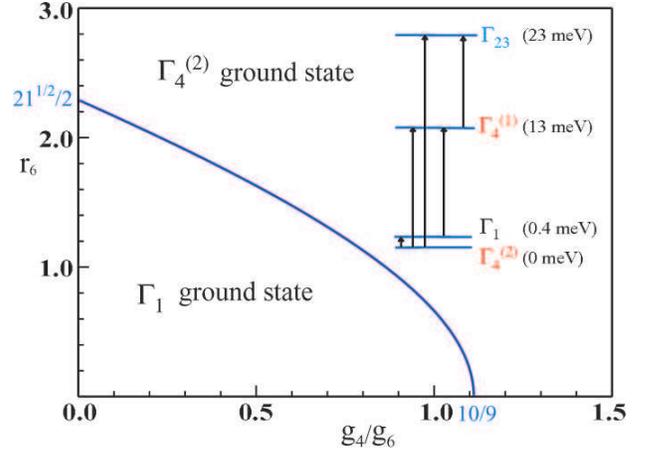}
\caption{\label{fig4} (Color online) The boundary between singlet and triplet ground states 
in skutterudites ($E_{1} = E_{4}^{(2)}$) as a function of $r_6$ and $g_4/g_6$, and 
the observed PrOs$_4$As$_{12}$ spectrum.
}
\end{figure}


Our determination of the CEF levels in PrOs$_4$As$_{12}$ 
reveals the reasons for the wide range of behaviours in 
different FS. The spectrum of CEF levels is largely determined by the $O_h$ symmetry field
of the eight nearest neighbor ions, and for Os the near equality of the
$\ell=4$ and $\ell=6$ strengths $g_4$ and $g_6$ implies nearly degenerate low-lying 
singlet (insulator) and triplet (AF) levels. The low temperature magnetic 
properties are determined by which of these phases happens to be the true ground state. 
In the CEF model, this is specified by the two parameters $g_4/g_6$ and $r_6$ (Fig. 7); 
in PrOs$_4$As$_{12}$, which has a triplet ground state, we estimate $g_4/g_6 \approx 1.15$ 
and $r_6 \approx 1.2$. The $T_h$ symmetry pnictogen CEF (proportional to $r_6$) 
acts to stabilize the triplet state, and can itself lead to a triplet ground state if 
$r_6$ is sufficiently large to cross the phase boundary shown in Fig. 7. 

In principle, one can extend our approach to calculate the ground states of 
other Pr-FS by determining its crystal structure and $g_4/g_6$ ratio. 
The necessity of using the $T_h$-symmetry of As 
rather than the $O_h$-symmetry pure Os form to explain the observed excitations
shows that the detailed pnictogen geometry is 
important in determining the CEF levels.
Indeed, the AF-ordered ground state in PrOs$_4$As$_{12}$ can arise from a 
$\Gamma_4^{(2)}$ triplet magnetic ground state, while the superconducting PrOs$_4$Sb$_{12}$
has nonmagnetic $\Gamma_1$ singlet ground state. 
The nearly degenerate first excited state $\Gamma_1$ at 0.4~meV ($\sim$ 4~K), 
and its temperature and field dependence (Figs.~4 and 6), 
may explain the presence of multiple transitions 
in the specific heat (in $C(T)/T$ versus $T$) and its 
field dependence \cite{Yuhasz,Maple06,ho}.  

\begin{center}
${\bf II. ~~SUMMARY}$
\end{center}

To understand the observed Pr$^{3+}$ CEF 
levels, one must incorporate the As ions' contribution  
to the CEF Hamiltonian \cite{Takegahara00,Takegahara01}, 
in addition to the usual Os cubic field terms.
A comparison of our CEF calculations using this more general Hamiltonian with our 
experimental results shows that the Pr$^{3+}$ CEF level scheme in 
PrOs$_4$As$_{12}$ consists of a $\Gamma_4^{(2)}$ magnetic triplet ground state,
a nearly degenerate $\Gamma_1$ singlet excitation, and higher 
$\Gamma_4^{(1)}$ magnetic triplet and $\Gamma_{23}$ nonmagnetic doublet 
excited states.
We find that contributions in the CEF Hamiltonian due to As are 
important in determining the neutron excitation 
intensities in PrOs$_4$As$_{12}$; our results 
differ qualitatively from the predictions of the conventional CEF Hamiltonian 
\cite{LLW,Birgeneau}, and therefore 
provide a microscopic understanding for its AF ground state.

\begin{center}
${\bf ACKNOWLEDGEMENT}$
\end{center}

We thank R. J. Birgeneau, B. C. Sales and D. Schultz for helpful discussion. 
The work at UT/ORNL was supported by the U.S. DOE under grant DE-FG02-05ER46202,
ORNL is managed by UT-Battelle, LLC, for the U.S. DOE under contract DE-AC05-00OR22725. 
Work at UCSD was supported by the U.S. DOE
under grant DE-FG02-04ER46105 and by the NSF under grant 
DMR-0335173. Work on SPINS was supported in part by the National Science Foundation under 
Agreement No. DMR-0454672.

\end{document}